# Imaging exciton-polariton transport in MoSe$_2$ waveguides

F. Hu[1,2], Y. Luan[1], M. E. Scott[3], J. Yan[4,5], D. G. Mandrus[4,5], X. Xu[3,6], Z. Fei[1,2]

[1]Department of Physics and Astronomy, Iowa State University, Ames, Iowa 50011, USA
[2]Division of Materials Sciences and Engineering, Ames Laboratory, U.S. DOE, Iowa State University, Ames, Iowa 50011, USA
[3]Department of Physics, University of Washington, Seattle, Washington 98195, USA
[4]Materials Science and Technology Division, Oak Ridge National Laboratory, Oak Ridge, Tennessee 37831, USA
[5]Department of Materials Science and Engineering, University of Tennessee, Knoxville, Tennessee 37996, USA
[6]Department of Materials Science and Engineering, University of Washington, Seattle, Washington 98195, USA

**The exciton polariton (EP), a half-light and half-matter quasiparticle, is potentially an important element for future photonic and quantum technologies[1-4]. It provides both strong light-matter interactions and long-distance propagation that is necessary for applications associated with energy or information transfer. Recently, strongly-coupled cavity EPs at room temperature have been demonstrated in van der Waals (vdW) materials due to their strongly-bound excitons[5-9]. Here we report a nano-optical imaging study of waveguide EPs in MoSe$_2$, a prototypical vdW semiconductor. The measured propagation length of the EPs is sensitive to the excitation photon energy and reaches over 12 μm. The polariton wavelength can be conveniently altered from 600 nm down to 300 nm by controlling the waveguide thickness. Furthermore, we found an intriguing mode back-bending dispersion close to the exciton resonance. The observed EPs in vdW semiconductors could be useful in future nanophotonic circuits operating in the near-infrared to visible spectral regions.**

In recent years, van der Waals (vdW) materials have emerged as a new material system supporting various types of polaritons with unique properties[10,11]. For example, graphene was discovered to support surface plasmon polaritons with high confinement, long lifetime and gate tunability[10-15]. Thin flakes of hexagonal boron nitride were proven to support hyperbolic phonon polaritons with wavelengths down to a few hundred nanometres[10,11,16-18]. These unique polaritons make both materials promising for nanophotonic applications in the terahertz to mid-infrared frequency regime. Group IVB transition-metal dichalcogenides (TMDs) with chemical formula MX$_2$ (M = Mo, W; X = S, Se) are vdW semiconductors with sizable bandgaps and strongly bounded excitons[19-22]. These excitons can couple with photons to form half-light and half-matter quasiparticles, namely exciton polaritons (EPs)[1-4]. Due to the large exciton binding energy, polaritons in TMDs are expected to be stable and robust at ambient conditions, thus suitable for technological applications. Indeed, far-field optical studies of TMDs embedded in micro-cavities captured the spectroscopic signatures of strongly-coupled cavity EPs[5-8]. Real-space characteristics (e.g. propagation, confinement, and interference) of the TMD polaritons, on the other hand, have not been addressed. Very recently, an imaging study of WSe$_2$ with the *aperture-type* scanning near-field optical microscope was reported[23], where interactions between waveguide photons and excitons were observed. Nevertheless, the characteristic dispersion relation of the waveguide EPs was not observed.

In this work, we performed nano-optical imaging studies of TMD planar waveguides, where EPs were formed due to the strong coupling between excitons and waveguide photons[24-28]. In order to probe these waveguide EPs, we used a *scattering-type* scanning near-field optical microscope (s-SNOM) that is built based on a tapping-mode atomic force microscope (AFM) with a sharp metallized tip (Fig. 1a). The spatial resolution of the s-SNOM defined by the radius of curvature of the tip apex is about 25 nm. In addition, the AFM tip is illuminated by a *p*-polarized laser beam. The high spatial resolution and *p*-polarized excitation of the s-SNOM enables effective dispersion mapping of individual TM waveguide modes. For signal detection, we use a concave mirror to collect photons back-scattered off the coupled tip-sample system (Fig. 1b) and these photons are counted by an amplified silicon photodetector. The samples studied here are exfoliated $MoSe_2$ thin flakes on standard $SiO_2$/Si wafers. In order to cover EPs due to the A excitons (~1.55 eV) of $MoSe_2$ (Supplementary Fig. S4), we used a continuous-wave Ti:$Al_2O_3$ laser that can be tunable from 1.3 to 1.8 eV.

In Fig. 1c-g, we show a selected dataset of s-SNOM images taken on a 156-nm-thick $MoSe_2$ planar waveguide, where we plot the normalized near-field amplitude (*s*) at various excitation laser energies (*E*). Within these images, we see clear interference fringes on $MoSe_2$ parallel to its edge (dashed lines), and these fringes demonstrate a clear energy dependence. First, we find that the fringe period increases systematically with decreasing *E*. In addition, the fringe intensity shows a significant enhancement at lower *E*. Moreover, we notice that the fringes extend further into the sample interior as *E* decreases. For example, at *E* = 1.35 eV (Fig. 1g and Supplementary Fig. S8), fringes can be seen 30 μm away from the sample edge.

Based on the above observations, we hypothesize that these fringes are generated due to the interference between photons collected by the detector from different paths. As illustrated in Fig. 1b, the collected photons come from two major paths. In the first path (marked with 'P1'), incident photons are scattered back directly by the s-SNOM tip. In the second path (marked with 'P2'), the laser-illuminated tip launches in-plane propagative modes inside the sample. As discussed in detail below, these in-plane modes correspond to the $TM_0$ waveguide modes. The waveguide modes propagate radially away from the tip and then get scattered into photons by the sample edge. Photons collected from paths 'P1' and 'P2' have a phase delay that scales with the distance between the tip and the sample edge. Therefore as the tip scans towards the edge of $MoSe_2$, one expects to see oscillations of photon intensity due to the interference of photons from the two photon paths. Other possible photon paths play less important roles as discussed in the Supplementary Information.

The above hypothesis implies a sensitive dependence of the fringes pattern on the orientation of the sample edge relative to the incident beam. In the configuration described in Fig. 1b and Fig. 2a, the laser beam is in the *x-z* plane and the sample edge is along the *y* direction, so the laser beam is perpendicular to the sample edge (referred to as 'perpendicular configuration'). In this configuration, photons collected through path 'P2' (Fig. 2a) are mainly from waveguide modes (marked with 'w.m.') propagating along the -*x* direction (Supplementary Information). Therefore, the fringe period in the perpendicular configuration ($\rho_\perp$) is expected to be

$$\rho_\perp \approx \lambda_p \left[1 - (\lambda_p / \lambda_0) \cos\alpha \right]^{-1}, \quad (1)$$

where $\lambda_p$ is the wavelength of the waveguide mode, $\lambda_0$ is the excitation laser wavelength, and $\alpha \approx 30°$ is the incident angle of the laser beam relative to the *x-y* plane. In another configuration (Fig. 2b), the sample edge is along the *x* direction and thus the in-plane projection of the incident beam is parallel to the sample edge (referred to as 'parallel configuration'). Here, photons in path 'P2' are mainly those scattered from waveguide modes traveling in an angle of $\phi$ relative to the $y$ direction (Fig. 2b), where $\phi = \sin^{-1}[(\lambda_p/\lambda_0)\cos\alpha]$ obtained by matching the boundary condition (momentum conservation along the edge direction) (Supplementary Information). Therefore, the fringe period in the parallel configuration ($\rho_{//}$) is expected to be

$$\rho_{//} \approx \lambda_p \left[1/\cos\phi - (\lambda_p/\lambda_0)\tan\phi\cos\alpha\right]^{-1}. \quad (2)$$

Based on Eqs. 1 and 2, we know that $\rho_{//}$ is smaller than $\rho_\perp$. Therefore, edge orientation dependence study provides a convenient way to test our hypothesis about fringe formation.

Figures 2c,d show near-field amplitude images taken at $E = 1.38$ eV (corresponding to $\lambda_0 = 900$ nm) in the perpendicular and parallel configurations, respectively. Apparently, the fringes obtained in the parallel configuration (Fig. 2d) are denser than those in the perpendicular configuration (Fig. 2c). For the purpose of quantitative comparison, we extracted line profiles perpendicular to the fringes directly from Fig. 2c,d, and then performed Fourier Transform (FT) analysis on these fringe profiles to accurately determine the fringe periods. Thus-obtained fringe profiles are plotted in Fig. 2e,f and the corresponding FT profiles are given in Fig. 2g,h. For convenience, we set the horizontal axis of the FT profiles to be $1/\rho_{//}$ and $1/\rho_\perp$ for parallel and perpendicular configurations, respectively. Considering that both $\rho_{//}$ and $\rho_\perp$ are clearly smaller than 1 μm, we only pay attention to the FT peaks above 1 μm$^{-1}$. In this regime, we can locate a dominant FT peak at 1.64 μm$^{-1}$ for $1/\rho_\perp$ and 2.51 μm$^{-1}$ for $1/\rho_{//}$. Therefore we have $\rho_\perp = 610$ nm and $\rho_{//} = 398$ nm, based on which we can calculate $\lambda_p$ to be 383 nm and 377 nm for perpendicular and parallel configurations, respectively (Eqs. 1 and 2). The values of $\lambda_p$ acquired from the two configurations are highly consistent with a deviation less than 2%, which validates our hypothesis and analysis.

Following the above methodology, we can now analyze the s-SNOM imaging data taken at all other laser energies. Figures 3a plots the fringe profiles taken at excitation energies from 1.35 to 1.77 eV. Their corresponding FT profiles are shown in Fig. 3b, where we can locate the dominant peaks (marked with arrows) due to the waveguide mode. By accurately measuring the FT peak positions, we can extract $\rho_\perp$ and then calculate $\lambda_p$ with Eq. 1. In addition, the propagation length ($L_p$) of the waveguide mode can be estimated by measuring the linewidths of the FT peaks (Methods). Thus-obtained $L_p$, plotted in Fig. 3c as squares, is at least over 12 μm at low energies, currently limited by our device size (Supplementary Information). At higher energies close to or above the A exciton energy, $L_p$ drops rapidly to 2 μm or less. The general trend of the experimental $L_p$ is consistent with the theoretical estimation (solid curve in Fig. 3c) (Methods).

Based on the extracted $\lambda_p$ through fringe analyses, we construct the energy ($E$) - momentum ($q_p = 2\pi/\lambda_p$) dispersion relation of the waveguide mode in Fig. 1. The obtained experimental ($q_p$, $E$) data points (blue squares) are overlaid on top the calculated dispersion color map (Fig. 3d). As introduced in the Methods, the bright regions in the color maps represent various photonic/polaritonic modes existing in the sample/substrate system. For convenience, we use the free-space photon wavevector $k_0 = 2\pi/\lambda_0$ as the momentum unit, which leads to vertical dispersions of photons in air ($q = k_0$) and SiO$_2$ ($q = 1.46k_0$) marked by the green and blue dashed lines in Fig. 3d. Here in the dispersion map, we find a good agreement between experimental data points (squares) with a confined mode close to $q \approx 2.5k_0$ in the color map. According to mode analysis (Supplementary Fig. S5), this mode corresponds to the TM$_0$ waveguide mode inside MoSe$_2$.

To reveal the detailed features of the $TM_0$ mode, we show a zoomed-in view ($2.2k_0 < q < 2.8k_0$) of Fig. 3d in Fig. 3e, where a back-bending dispersion of the waveguide mode is clearly visualized near the A exciton energy (see also the dispersion data of the 110-nm-thick $MoSe_2$ sample in Supplementary Fig. S6). Such an 'anomalous' dispersion is in fact the characteristic behaviour of the EPs under measurements with fixed excitation energies (imaging experiments with a continuous-wave laser in our case). The commonly-accepted anti-crossing dispersion of the EPs, on the other hand, can be obtained by measurements at fixed momenta (e.g. spectroscopic studies of cavity polaritons at fixed incident angles)[1-8]. The fixed-energy imaging measurements intend to determine the polariton momenta ($q_p$) by searching horizontally the dispersion map (e.g. along horizontal dashed lines in Fig. 3f), while the fixed-momentum spectroscopic experiments are to locate the polariton energy ($E_p$) by sweeping vertically the dispersion map (e.g. along the vertical dashed lines in Fig. 3g). With both methods, one can obtain a series of ($q_p$, $E_p$) data points (blue crosses in Fig. 3f,g). The polariton dispersion reflected by these data points demonstrates either back-bending (Fig. 3f) or anti-crossing (Fig. 3g) features. Note that the back-bending dispersion also suggests that polaritons are subject to broadening, which introduces finite photonic spectral weight at the gap between the top and bottom polaritonic branches. The broadening is mainly due to scatterings of EPs with longitudinal optical phonons that can be strongly suppressed at cryogenic temperature (Supplementary Fig. S7). Back-bending dispersion has been observed previously in both plasmon and phonon polaritons[29,30]. Our experiment proves that the EPs also share this phenomenon. Based on the back-bended dispersion data points, we estimate a Rabi splitting energy ($E_{Rabi}$) of ~100 meV (yellow arrow in Fig. 3e), indicating a strong coupling between excitons and waveguide photons. The $E_{Rabi}$ value measured here in bulk $MoSe_2$ appear to be larger than those of atomic layers of TMDs[5-8] and smaller than that of bulk $WS_2$[9].

Finally, we performed s-SNOM imaging of $MoSe_2$ waveguides with different thicknesses. As shown in Fig. 4a,b, we plot the near-field images of two additional $MoSe_2$ waveguides with thicknesses of 77 and 110 nm, respectively. They are both taken at $E = 1.48$ eV in the perpendicular configuration. As described above, we determined the fringe period ($\rho_\perp$) by extracting fringe profiles (Fig. 4c) by FT analysis (Fig. 4d). Employing Eq. 1, we obtained $\lambda_p$ versus waveguide thickness (Fig. 4e), which shows good consistency with theory (black curve) (Supplementary Material). From Fig. 4e, we found that $\lambda_p$ can be altered from 600 to 300 nm by controlling the waveguide thickness. Note that the observed $TM_0$ polariton mode is cut off in $MoSe_2$ waveguides with a thickness less than ~70 nm (Fig. 4e). In order to explore polaritons in thinner flakes or even atomic layers of TMDs, other type of waveguide modes (e.g. $TE_0$ mode[23]) or other coupling methods (e.g. micro-cavity coupling[1-8]) have to be adopted.

By combining the s-SNOM technique with rigorous theoretical analyses, we uncovered the real-space characteristics of EPs in $MoSe_2$ waveguides. The observed polaritons have shown a small wavelength (down to 300 nm) and a long propagation length (up to 12 μm or above) under ambient conditions. These characteristics observed in our first generation devices are comparable to or even better than surface plasmon polaritons in graphene[12-15] and hyperbolic phonon polaritons in hexagonal boron nitride[16-18]. Through careful design and engineering, the TMD waveguides with tailored polaritonic modes could potentially be applied in miniaturized nanophotonic circuits for information or energy transfer in the near-infrared to visible regions. In addition, it will be interesting to perform polariton nano-imaging at cryogenic temperatures, where one could possibly visualize EPs with stronger coupling strength and longer propagation length (Supplementary Fig. S7). Future studies are also promising to explore new polaritonic characteristics and functionalities by patterning TMD flakes into nano-resonators or other types of photonic structures (e.g. photonic crystals). Our work opens up new avenues for studies of EPs and paves the way for future applications of TMDs in optoelectronics and nanophotonics.

**References**
1. Weisbuch, C. et al. Observation of the coupled exciton–photon mode splitting in a semiconductor quantum microcavity. *Phys. Rev. Lett.* **69**, 3314–3317 (1992).
2. Gibbs, H.M., Khitrova, G. & Koch, S.W. Exciton-polariton light-semiconductor coupling effects. *Nature Photon.* **5**, 275-282 (2011).
3. Tassone, F., Bassani, F. & Andreani, L.C. Quantum-well reflectivity and exciton-polariton dispersion. *Phys. Rev. B* **45**, 6023-6030 (1992).
4. Deng, H., Haug, H. & Yamamoto, Y. Exciton-polariton Bose-Einstein condensation. *Rev. Mod. Phys.* **82**, 1489-1537 (2010).
5. Liu, X. et al. Strong light-matter coupling in two-dimensional atomic crystals. *Nature Photon.* **9**, 30-34 (2015).
6. Dufferwiel, S. et al. Exciton–polaritons in van der Waals heterostructures embedded in tunable microcavities. *Nature Commun.* **6**, 8579 (2015).
7. Lundt, N. et al. Room-temperature Tamm-plasmon exciton-polaritons with a $WSe_2$ monolayer. *Nature Commun.* **7**, 13328 (2016).
8. Flatten, L.C. et al. Room-temperature exciton polaritons with two-dimensional $WS_2$. *Sci. Rep.* **6**, 33134 (2016).
9. Wang, Q. et al. Direct observation of strong light-exciton coupling in thin $WS_2$ flakes. *Opt. Express* 24, 7151-7157 (2016).
10. Basov, D.N., Fogler, M.M. & García de Abajo, F.J. Polaritons in van der Waals materials. *Science* **354**, 195 (2016).
11. Low, T. et al. Polaritons in layered two-dimensional materials. *Nature Mater.* **16**, 182-194 (2017).
12. Ju, L. et al. Graphene plasmonics for tunable terahertz metamaterials. *Nature Nanotechnol.* **6**, 630-634 (2011).
13. Fei, Z. et al. Gate-tuning of graphene plasmons revealed by infrared nano-imaging. *Nature* **487**, 82-85 (2012).
14. Chen, J. et al. Optical nano-imaging of gate-tunable graphene plasmons. *Nature* **487**, 77-81 (2012).
15. Woessner, G. et al. Highly confined low-loss plasmons in graphene-boron nitride heterostructures. *Nature Mater.* **14**, 421–425 (2015).
16. Dai, S. et al. Tunable phonon polaritons in atomically thin van der Waals crystals of boron nitride. *Science* **343**, 1125-1129 (2014).
17. Li, P. et al. Hyperbolic phonon-polaritons in boron nitride for near-field optical imaging and focusing. *Nature Commun.* **6**, 7507 (2015).
18. Yoxall, E. et al. Direct observation of ultraslow hyperbolic polariton propagation with negative phase velocity. *Nature Photon.* **9**, 674-678 (2015).
19. Splendiani, A. et al. Emerging photoluminescence in monolayer $MoS_2$. *Nano Lett.* **10**, 1271-1275 (2010).
20. Mak, K.F., Lee, C., Hone, J., Shan, J. & Heinz, T.F. Atomically thin $MoS_2$: a new direct-gap semiconductor. *Phys. Rev. Lett.* **105**, 136805 (2010).
21. Radisavljevic, B., Radenovic, A., Brivio, J., Giacometti & Kis, V.A. Single-layer $MoS_2$ transistors. *Nature Nanotechnol.* **6**, 147-150 (2011).

22. Wang, Q.H., Kalantar-Zadeh, K., Kis, A., Coleman, J.N. & Strano, M.S. Electronic and optoelectronics of two-dimensional transition metal dichalcogenides. *Nature Nanotechnol.* **7**, 699-712 (2012).
23. Fei, Z. et al. Nano-optical imaging of $WSe_2$ waveguide modes revealing light-exciton interactions. *Phys. Rev. B* **94**, 081402(R) (2016).
24. Katsuyama, T. & Ogawa, K. Excitonic polaritons in quantum-confined systems and applications to optoelectronic devices. *J. Appl. Phys.* **75**, 7607-7625 (1994).
25. van Vugt, L.K. et al. Exciton polaritons confined in a ZnO nanowire cavity. *Phys. Rev. Lett.* **97**, 147401 (2006).
26. Takazawa, K. et al. Fraction of a millimeter propagation of exciton polaritons in photoexcited nanofibers of organic dye. *Phys. Rev. Lett.* **105**, 067401 (2010).
27. Liscidini, M. et al. Guided Bloch surface wave polaritons. *Appl. Phys. Lett.* **98**, 121118 (2011).
28. Ellenbogen, T., Steinvurzel, P. & Crozier, K.B. Strong coupling between excitons in J-aggregates and waveguide modes in thin polymer films. *Appl. Phys. Lett.* **98**, 261103 (2011).
29. Arakawa, E.T., Williams, M.W., Hamm, R.N. & Ritchie, R.H. Effect of damping on surface plasmon dispersion. *Phys. Rev. Lett.* **31**, 1127-1129 (1973).
30. Schuller, E., Falge, H.J. & Borstel, G. Dispersion curves of surface phonon-polaritons with backbending. *Phys. Lett.* A **54**, 317-318 (1975).

**Acknowledgements**
F.H., Y.L. and Z.F. acknowledge startup support from Iowa State University and Ames Laboratory. The nano-optical imaging setup was partially supported by the W. M. Keck foundation. The work at UW was supported by the U.S. DOE Basic Energy Sciences, Materials Sciences and Engineering Division (DE-SC0008145 and SC0012509). The work at ORNL (JQY and DGM) was supported by the U.S. Department of Energy, Office of Science, Basic Energy Sciences, Materials Sciences and Engineering Division.

**Author contributions**
Z.F. conceived the ideas and designed the experiments. F.H. carried out the s-SNOM experiments and collected the data. Z.F., F.H. and Y.L. performed theoretical analyses and modeling of the data. X.X., M.E.S., J.Y. and D.G.M. synthesized the $MoSe_2$ crystals and fabricated the waveguide devices. Z.F., X.X., F.H. and Y.L. wrote the paper.

**Additional information**
The authors declare no competing financial interests. Reprints and permission information is available online at http://npg.nature.com/reprintsandpermissions. Correspondence and requests for materials should be addressed to Z.F. (zfei@iastate.edu).

**Figure Legends**

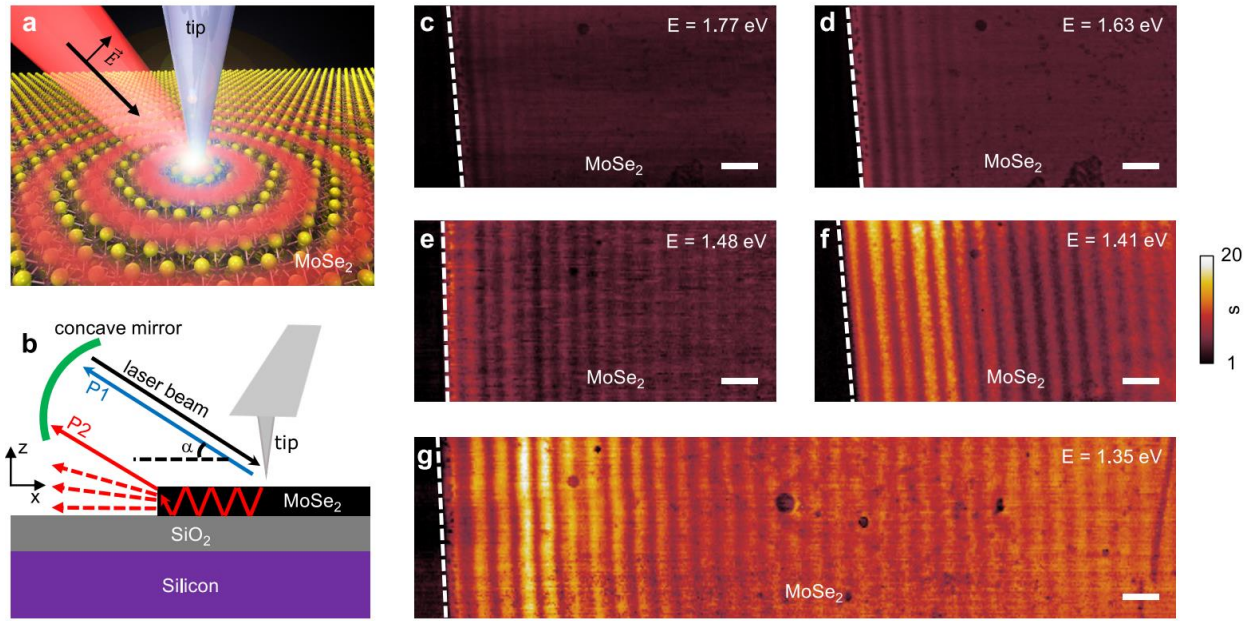

**Figure 1. Nano-optical imaging of a MoSe₂ planar waveguide. a**, Schematics of concentric waveguide modes in MoSe₂ launched by the laser-illuminated s-SNOM tip. **b**, Illustration of the experimental setup where the incident beam is aligned perpendicular to the sample edge that is along the *y* direction. We also sketch here the two major paths ('P1' and 'P2') where photons can be collected by the concave mirror. **c**-**g**, Selected s-SNOM imaging data of a 156-nm-thick MoSe₂ planar waveguide taken at various laser energies (*E*). Here we plot the near-field amplitude (*s*) normalized to that of the SiO₂/Si substrate. The dashed lines mark the sample edge. The scale bars represent 1 μm.

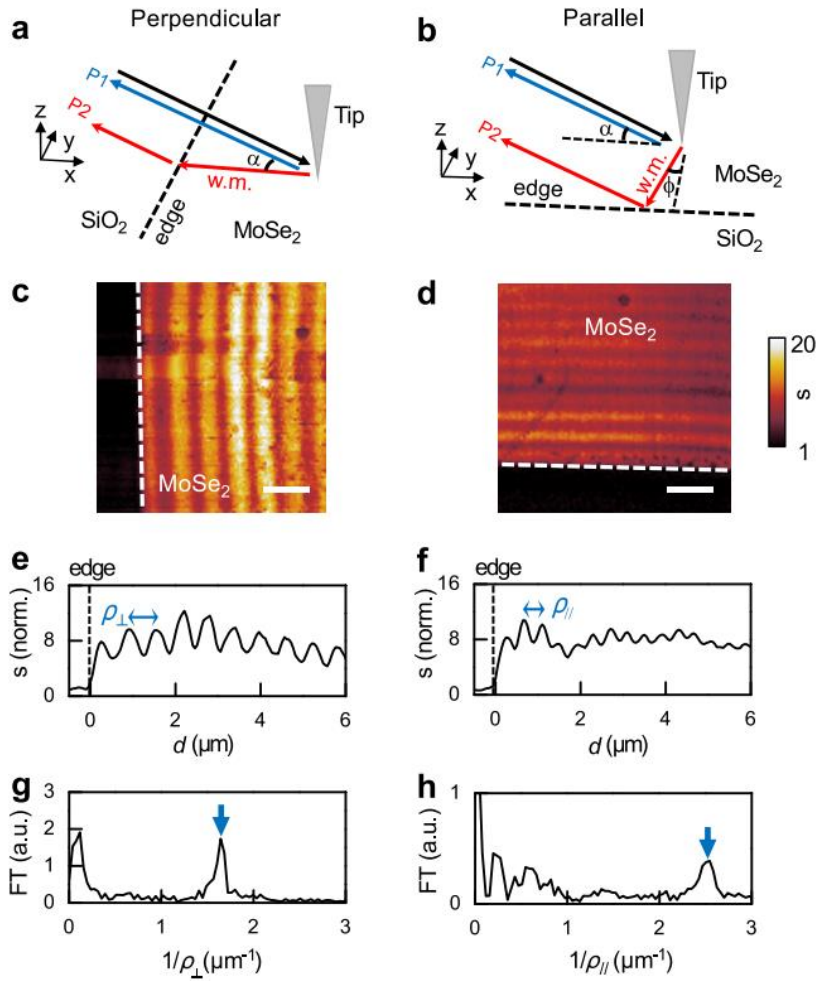

**Figure 2. Edge-orientation dependence study. a**, Illustration of the perpendicular configuration, where the incident beam (black arrow) is perpendicular to the sample edge. **b**, Illustration of the parallel configuration, where the *x-y* plane projection of the incident beam is parallel to the sample edge. The labeling 'w.m.' in **a** and **b** represents waveguide modes. **c,d**, Near-field amplitude images of MoSe$_2$ taken at $E$ = 1.38 eV in the perpendicular and parallel configurations, respectively. The dashed lines mark the sample edge. The scale bars represent 1 μm. **e,f**, Real-space line profiles extracted perpendicular to the fringes in **c** and **d**, respectively. Here *d* is the distance between the tip and the sample edge. **g,h**, Fourier transform (FT) analysis of the real-space profiles in **e** and **f**, respectively.

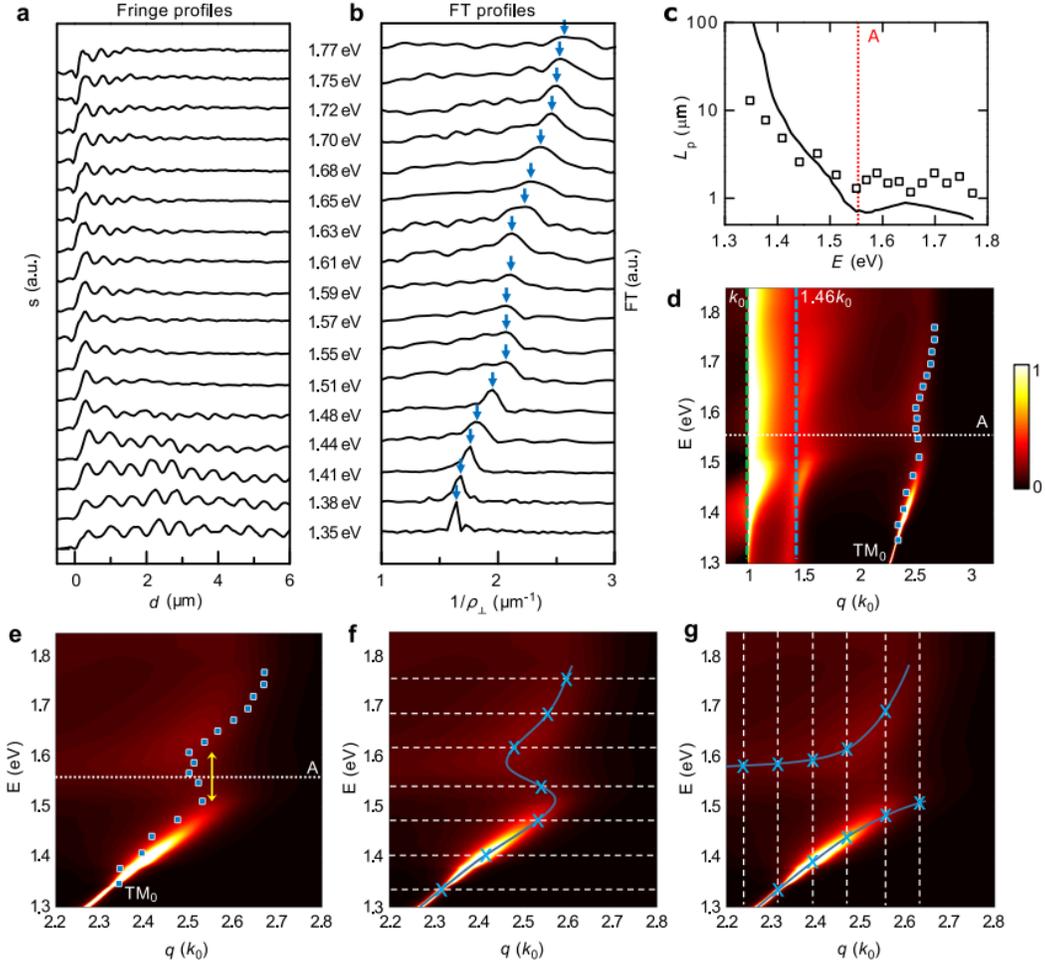

**Figure 3. Dispersion analysis. a,b**, Real-space fringe profiles and the corresponding FT profiles of the 156-nm-thick MoSe$_2$ waveguide taken at various excitation energies (1.35 - 1.77 eV) in the perpendicular configuration. All the profiles are displaced vertically for clarity. The arrows in **b** mark the FT peak associated with the measured waveguide mode in MoSe$_2$. **c**, Propagation length ($L_p$) of the measured waveguide mode from both experiment (squares) and theory (curve). **d**, Experimental dispersion data points (blue squares) overlaid on the calculated dispersion color map. **e**, A zoomed-in version of panel **d** at the $q$ range from $2.2k_0$ to $2.8k_0$. The yellow arrow marks the Rabi splitting energy. **f**, Illustration of the fixed-$E$ experiments with horizontal line cuts across the dispersion map. **g**, Illustration of the fixed-$q$ experiments with vertical line cuts across the dispersion map. The blue crosses in **f** and **g** mark the positions with maximum photonic spectral weight along the line cuts. The color maps in **d**-**g** plot the imaginary part the $p$-polarized reflection coefficient Im($r_p$) (Methods).

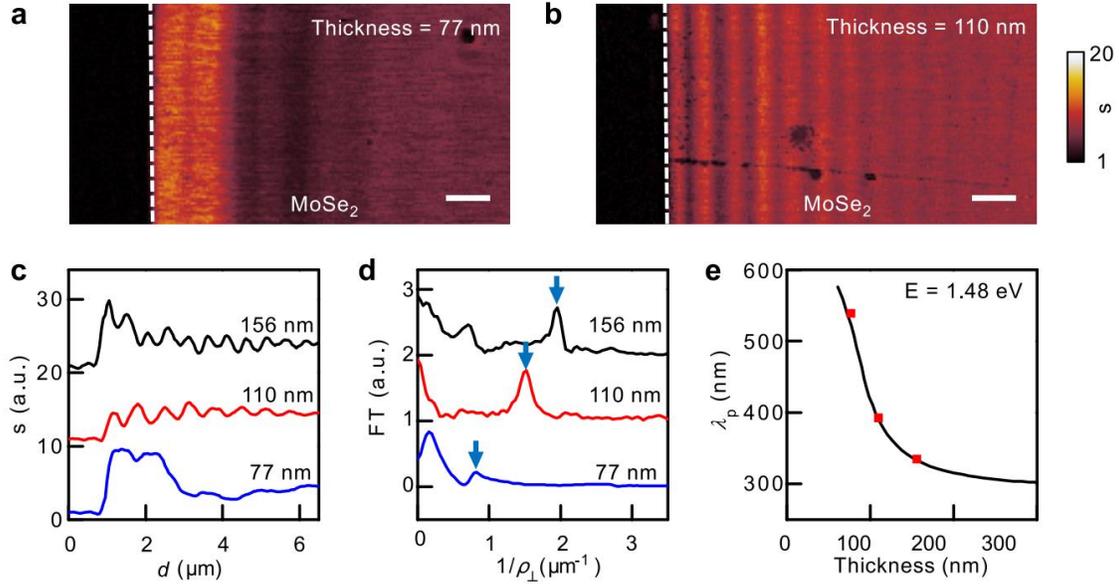

**Figure 4. Waveguide-thickness dependence study. a,b**, Near-field imaging data of MoSe$_2$ planar waveguides with thicknesses of 77 and 110 nm, respectively. These images were taken at $E = 1.48$ eV in the perpendicular configuration. The scale bars represent 1 μm. **c,d**, Real-space fringe profiles and the corresponding FT profiles of MoSe$_2$ with various thicknesses. **e**, Thickness dependence of polariton wavelengths ($\lambda_p$) from both experiment (data points) and theory (black curve) at $E = 1.48$ eV.

## Methods

**Experimental setup.** For nano-optical imaging experiments, we used a scattering-type scanning near-field optical microscope (s-SNOM, Neaspec) that is built based on a tapping-mode atomic force microscope (AFM) operating with a tapping frequency of about 270 kHz and a tapping amplitude of about 50 nm. A pseudo-heterodyne interferometric detection module is implemented in our s-SNOM to extract both the scattering amplitude (*s*) and phase ($\psi$) of the near-field signal. In this work, we discuss mainly the *s* signal that is sufficient for describing all the characteristics of the EPs. In order to subtract the background signal, we demodulated the near-field signal at the 3$^{rd}$ harmonics of the tapping frequency of the AFM tip. In all the displayed near-field images, we plotted *s* normalized to that of the SiO$_2$/Si substrate. For optical excitation, we used a Ti:Al$_2$O$_3$ laser operating in the continuous-wave mode. The photon energy of the Ti:Al$_2$O$_3$ laser can be conveniently tunable from 1.3 to 1.8 eV. The samples studied here are MoSe$_2$ planar waveguides fabricated using the mechanical exfoliation method. The substrates used for these samples are standard Si wafers with a 300-nm-thick thermal oxide layer on the top. Our s-SNOM experiments were all performed at ambient conditions.

**Dispersion calculation and analysis.** The dispersion color maps shown in Fig. 3d-g were obtained by evaluating numerically the imaginary part of the reflection coefficients Im($r_p$) of the multilayer sample/substrate system. The bright curves shown in the color map correspond to various photonic and polaritonic modes in the entire system. Considering that the electric field right underneath the s-SNOM probe is perpendicular to the sample surface, only transverse magnetic (TM) waveguide modes are excited. Therefore, we only consider *p*-polarized reflection coefficient $r_p(q, E)$ in the dispersion calculation. By using the transfer matrix method, we numerically calculate Im($r_p$) of the entire air/MoSe$_2$/SiO$_2$/Si system. The photonic/polaritonic modes appear at the ($q$, $E$) positions where Im($r_p$) diverges or maximizes. Therefore, we can conveniently estimate numerically the mode wavelength ($\lambda_p = 2\pi/q_p$) through the calculated dispersion color maps. In order to fit the experimental dispersion data points (Fig. 3d), we adopted the experimental *ab*-plane dielectric constants of MoSe$_2$ from previous optical measurement (Supplementary S4). The *c*-axis dielectric constant ($\varepsilon_c$) of MoSe$_2$ is a fitting parameter, which was set to be 8.3 throughout the entire spectral range of our experiment (1.3 – 1.8 eV). The good agreement between experimental data points and calculated dispersion plot validates such an assumption.

**Determining the propagation length of the EPs.** In order to determine quantitatively the propagation length ($L_p$) of the measured waveguide EPs, we first extract the linewidths (plotted in Supplementary Fig. S9a) of the FT peaks of the waveguide mode (Fig. 3b) and then determine $L_p$ using Eq. S11 in the Supplementary Information. Thus-obtained $L_p$ data points are plotted in Fig. 3c. Note that $L_p$ at the lowest energy ($E = 1.35$ eV) is most likely underestimated due to limited resolution of the FT profiles originated from the finite size (~ 30 μm) of our device (Supplementary Information). In order to estimate theoretically $L_p$, we approximate the tip-launched waveguide mode as cylinder waves with a wave function of $Ax^{-1/2}e^{i(qx-\omega t)}$. Here, $q = q_1 + iq_2$ is the complex in-plane momentum of the waveguide mode. Therefore, the amplitude of the wave decays as: $Ax^{-1/2}e^{-x/(2L_p)}$, where the propagation length $L_p$ equals to $(2q_2)^{-1}$. For an anisotropic material like MoSe$_2$ with an *ab*-plane dielectric function of $\varepsilon_{ab} = \varepsilon_1 + i\varepsilon_2$ (Supplementary Fig. S4) and an *c*-axis dielectric constant of $\varepsilon_c \approx 8.3$, we have $q = \sqrt{\varepsilon_c k_0^2 - (\varepsilon_c/\varepsilon_{ab})k_z^2}$. Based on this equation, we have an analytic formula of $L_p = [\varepsilon_2/(2\varepsilon_1 q_1)]\left[1 - \sqrt{1-(\varepsilon_2^2/\varepsilon_1^2)(\varepsilon_c k_0^2/q_1^2 - 1)}\right]^{-1}$. By adopting the $q_1 = 2\pi/\lambda_p$ data of the TM$_0$ waveguide mode (Fig. 3), we can calculate $L_p$ using the above formula. The calculated result is plotted in Fig. 3c as the solid curve. The general trend of the theoretical curve is consistent with the experimental data (squares in Fig. 3c).

**Data availability.** The data that support the plots within this paper and other findings of this study are available from the corresponding author upon reasonable request.

<div style="text-align:center">

**Supplementary Information for**
**"Imaging exciton-polariton transport in MoSe$_2$ waveguides"**

</div>

F. Hu[1,2], Y. Luan[1], M. E. Scott[3], J. Yan[4,5], D. G. Mandrus[4,5], X. Xu[3,6], Z. Fei[1,2]*

[1]Department of Physics and Astronomy, Iowa State University, Ames, Iowa 50011, USA
[2]Division of Materials Sciences and Engineering, Ames Laboratory, U.S. DOE, Iowa State University, Ames, Iowa 50011, USA
[3]Department of Physics, University of Washington, Seattle, Washington 98195, USA
[4]Materials Science and Technology Division, Oak Ridge National Laboratory, Oak Ridge, Tennessee 37831, USA
[5]Department of Materials Science and Engineering, University of Tennessee, Knoxville, Tennessee 37996, USA
[6]Department of Materials Science and Engineering, University of Washington, Seattle, Washington 98195, USA

*Correspondence to: zfei@iastate.edu.

**List of contents**

**1. Discussion and analysis about fringe formation**
    **1.1 Photon path 'P3'**
    In the main text, we discuss two major photon paths ('P1' and 'P2') for both perpendicular and parallel configurations (Fig. 1, Fig. 2 and Supplementary Fig. S1). In path 'P1', photons are scattered directly by the tip back to the detector. In path 'P2', photons first transfer into propagative waveguide exciton polaritons (EPs) and then scatter to photons when reaching the edge of the sample. In addition to the two photon paths, there is another possible photon path 'P3' that could contribute to the signal collection process. In photon path 'P3', the photons first transfer into EPs and then reflected backward to the tip after reaching the sample edge. The reflected EP modes can be scattered back to detector by the AFM tip. This path has been extensively discussed in previous nano-infrared imaging studies of graphene plasmons[1] and hexagonal boron nitride (hBN) polaritons[2]. In the current work, 'P3' plays a less significant role compared to 'P2' for the following two reasons. First, the momenta ($q$) of the modes involved in our experiments is much closer to the free-space photon wavevector ($k_0$). As a result, only a small portion of the EP modes are reflected back by the sample edge. In addition, the round-trip propagation of the EP modes in

'P3' also suffers more damping compared to 'P2' due to the longer traveling distance. As a result, we didn't see clear experimental evidence of fringes due to the interference between photons from paths 'P1' and 'P3'.

### 1.2 Edge excitation

The photon paths 'P2' and 'P3' discussed above all assume the SNOM tip as the only launcher of the EPs. In fact, the sample edge can also launch polaritons when illuminated by the laser beam. Nevertheless, due to the small size of the focused laser beam (~ 2 µm), edge launching is only possible when the tip-edge distance ($d$) is very small ($d < 1$ µm). Therefore, the fringes that are over 1 µm away from the edges are solely due to the effects of tip scattering or launching ('P1' and 'P2'). When performing FT analysis (Fig. 3b) on the fringe profiles (Fig. 3a), we cut off a short part of the fringe profiles (~ 1 µm close to the edge) in order to avoid the complications involving edge launching processes.

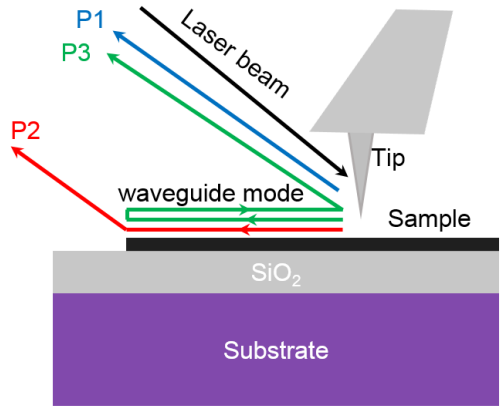

**Supplementary Figure S1.** Illustration of various paths from which the photons can be collected by the detectors.

### 1.3 Perpendicular configuration

In Fig. 2a and related text in the manuscript, we discuss the perpendicular configuration of the experimental setup, where the laser beam is perpendicular to the edge of sample. Based on this configuration, we obtain an equation (Eq. 1 in the main text) that describes the relation between the fringe period ($\rho_\perp$) and the wavelength of the EPs ($\lambda_p$). Now we introduce in detail how this equation is derived. As discussed above, the fringes observed in our experiments are mainly due to the interference between collected photons from paths 'P1' and 'P2'. In the latter path ('P2'), the tip-launched EPs propagate radially away from the tip, so they could in principle be scattered into free-space photons from any positions along the sample edge. Nevertheless, the concave mirror with a relatively small collection angle of less than 14° (beam size < 1 inch, focusing length ≈ 2 inches) collects mainly the scattered photons from waveguide EPs propagating along the direction perpendicular to the edge.

In order to elucidate that, we illustrate in Supplementary Fig. S2 the process of edge scattering where the incident laser beam is in the *x-z* plane and the sample edge is along the *y* direction (perpendicular configuration). Here we assume, tip-launched EPs with a momentum of $q_p$ propagate to the edge with a random angle ($\phi$) with respect to the *x*-axis. At the edge, the EPs are scattered into free-space photons with a wavevector of $k_0$. The angle between the scattered photons and the optical axis of the concave mirror is $\theta$. As discussed above, $\theta$ should be within the collection angle of the concave mirror: $\theta < 14°$. Therefore, photon wavevector along the y direction ($k_y$) should satisfy $k_y < k_0\sin(\theta) < k_0\sin(14°) = 0.242k_0$. In the edge scattering process, momentum is conserved along the edge direction (y direction), so we have $k_y = q_p\sin(\phi) < 0.242k_0$. Considering that the EPs of $MoSe_2$ studied in the current work are more confined in space than photon modes in $SiO_2$: $q_p > 1.46k_0$, we have $\phi < 9.8°$. Since the above upper bound limit analysis is far from stringent, the actual average $\phi$ will be much closer to zero with a more careful analysis. Furthermore, EPs with finite $\phi$ will be subjected to higher propagation loss and larger reflection coefficient than those with normal incidence towards the edge ($\phi = 0$). Therefore, we assume $\theta = \phi = 0°$ in the following analysis. The error bar of the extracted $\lambda_p$ based on such an assumption is less than $1 - \cos(9.8°) \approx 1.4\%$.

Under the assumption of $\theta = \phi = 0°$, we estimate the phase difference ($\Phi$) between photons collected from paths 'P1' and 'P2' to be:

$$\Phi = q_p d - k_0 d \cos\alpha = 2\pi d\left(1/\lambda_p - \cos\alpha/\lambda_0\right), \tag{S1}$$

where $d$ is the distance between the tip and the sample edge, $\lambda_p$ and $\lambda_0$ are the wavelengths of waveguide EPs and free-space photons, respectively. If $d$ changes by a distance that equals to the fringe period ($\rho_\perp$), $\Phi$ will change by $2\pi$. Therefore we have

$$\rho_\perp = \lambda_p\left[1 - \left(\lambda_p/\lambda_0\right)\cos\alpha\right]^{-1}. \tag{S2}$$

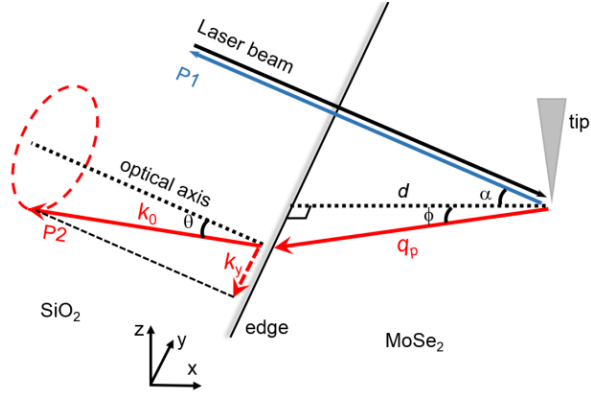

**Supplementary Figure S2.** Illustration of the perpendicular configuration of the nano-optical experimental setup.

### 1.4 Parallel configuration

Figure S3 illustrates the parallel configuration of the experimental setup, where the incident laser beam is kept in the x-z plane and the sample edge is along the x direction. Similar to the perpendicular configuration, the fringes are formed due to the phase difference between photons collected by the detector from paths 'P1' and 'P2'. In path 'P2', tip-launched EPs with an in-plane momentum of $q_p$ propagate toward the edge with an angle of $\phi$ with respect to the y axis and get scattered into free-space photons ($k_0$) when reaching the sample edge. Again, the angle $\theta$ between scattered photons and the optical axis should be less than the collection angle of the concave mirror: $\theta < 14°$. From Supplementary Fig. S3, we find that the x component of the wavevector of the scattered photons ($k_x$) should satisfy: $k_0\cos(\alpha + 14°) < k_x < k_0\cos(\alpha - 14°)$, where $\alpha = 30°$ is the angle between the optical axis of the concave mirror and the x-y plane. Therefore, we have $0.72k_0$

$< k_x < 0.96k_0$. Considering the momentum conservation along the direction of the scattering edge (x direction): $k_x = q_p\sin(\phi)$, so we have $0.72k_0/q_p < \sin(\phi) < 0.96k_0/q_p$. Based on this inequality, we have $16.8° < \phi < 22.6°$ in the case of $q_p \sim 2.5k_0$ (e.g. $TM_0$ mode in the 156-nm-thick $MoSe_2$ sample). The angle deviation is quite small, so we assume $\theta = 0$ for convenience. In this case, $k_x = q_p\sin(\phi) = k_0\cos(\alpha)$, $\phi = 20.3°$ when $q_p \sim 2.5k_0$. The uncertainty of the estimated $\lambda_p$ due to such an assumption is estimated to be less than $\cos(16.8°) - \cos(20.3°) \approx 1.9\%$.

Under the assumption of $\theta = 0$, we estimate the phase difference ($\Phi$) between photons collected from paths 'P1' and 'P2' to be:

$$\Phi = q_p d/\cos\phi - k_0 d \tan\phi\cos\alpha = 2\pi d\left(1/\cos\phi/\lambda_p - \tan\phi\cos\alpha/\lambda_0\right) \tag{S3}$$

where $\phi = \sin^{-1}[(\lambda_p/\lambda_0)\cos\alpha]$ obtained from $k_x = q_p\sin(\phi)$. If $d$ changes by a distance that equals to the fringe period ($\rho_{//}$), $\Phi$ will change by $2\pi$. Therefore we have

$$\rho_{//} \approx \lambda_p\left[1/\cos\phi - (\lambda_p/\lambda_0)\tan\phi\cos\alpha\right]^{-1} \tag{S4}$$

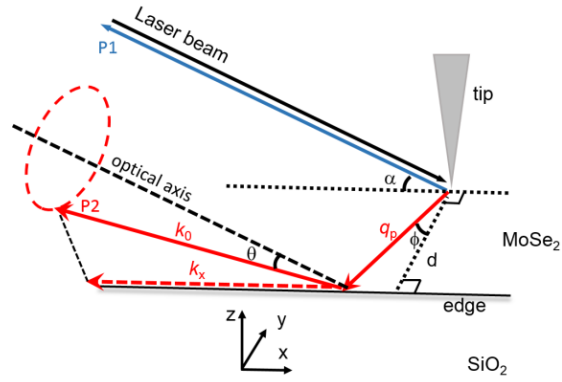

**Supplementary Figure S3.** Illustration of the parallel configuration of the nano-optical experimental setup.

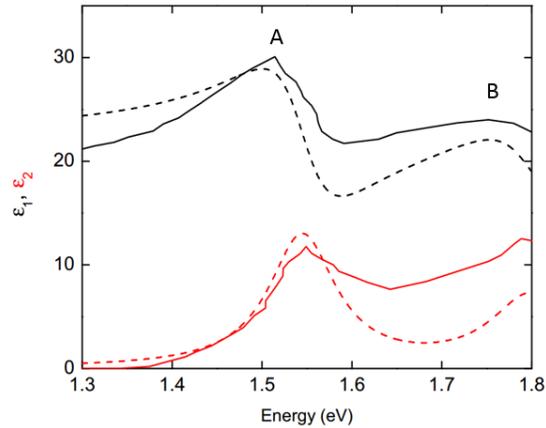

**Supplementary Figure S4.** In-plane (*ab*-plane) dielectric function of $MoSe_2$ for calculations of the EPs. The black and red curves are real and imaginary parts of the dielectric function, respectively. The solid lines are experimental data adopted from Ref. 3, which were measured in bulk $MoSe_2$ at room temperature. The dashed lines are modeled results that we constructed by using single Lorentzian oscillator for both A and B excitons (see detailed discussions in Section 4 of the Supplementary Information). The *c*-axis permittivity is set to be 8.3 throughout our energy region that produces a good fit to experimental dispersion data points (Fig. 3d in the main text).

## 2. Field-distribution calculation confirming TM$_0$ waveguide modes

In order to confirm the nature of the measured waveguide EPs, we performed field distribution calculations by matching boundary conditions in Maxwell's wave equations. The waveguide structure is plotted in Supplementary Fig. S5a. The calculations were performed considering 156-nm-thick MoSe$_2$ excited by a laser beam with a photon energy of 1.38 eV, but the general results apply also to other thicknesses (110 nm and 77 nm) or other laser energies. The ab-plane optical constants for MoSe$_2$ used in the calculation are adopted from Ref. 3 (Supplementary Fig. S4) and the *c*-axis permittivity is set to be 8.3 that is determined through dispersion fitting (Fig. 3d in the main text). The results are shown in Supplementary Fig. S5b and S5c, where we plot the *y*-component magnetic field ($H_y$) and *z*-component electric field ($E_z$) at various *z* positions, respectively. Note that the waveguide EPs is set to be propagating along the -*x* direction. From the field distribution, we know that the measured waveguide EPs correspond to TM$_0$ mode.

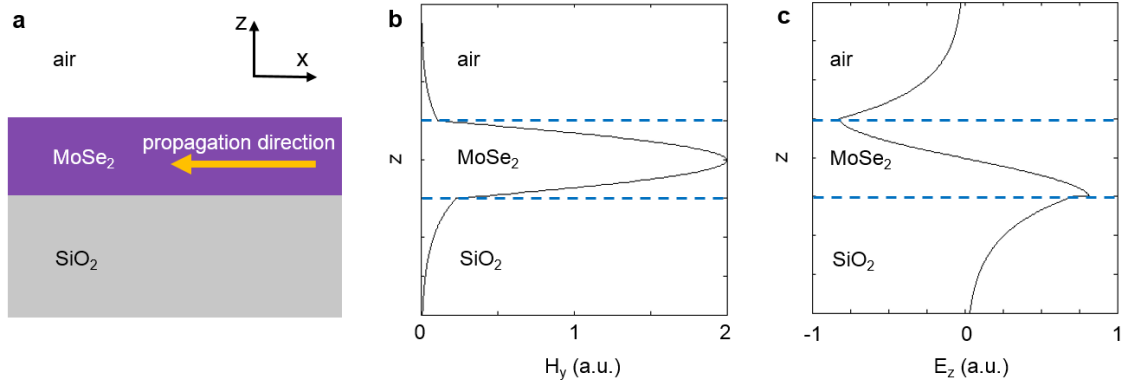

**Supplementary Figure S5. Field-distribution calculation confirming TM$_0$ waveguide modes.** **a**, Illustration of the MoSe$_2$ waveguide sandwiched by air and SiO$_2$. The waveguide EPs are propagating along the -*x* direction. **b,c**, Calculated $H_y(z)$ and $E_z(z)$ of the waveguide EPs in the 156-nm-thick MoSe$_2$ sample (Figs. 1 and 2 in the main text). The blue dashed lines here mark the top and bottom surfaces of the MoSe$_2$ layer.

## 3. Polariton dispersion of the 110-nm-thick MoSe$_2$ sample

In Fig. 3d,e of the main text, the dispersion data points (blue squares) were obtained from near-field amplitude images of the 156-nm-thick MoSe$_2$ waveguide. Here in Supplementary Fig. S6, we show additional dispersion data (blue squares) measured from the 110-nm-thick MoSe$_2$ waveguide. Again, these data points are in good agreement with the theoretical dispersion color map and they also show the back-bending feature close to the A exciton energy. We also notice that the polariton mode of the 110-nm-thick MoSe$_2$ sample shift to the low momentum region (~ 2$k_0$) compared to that of the 156-nm-thick one (~ 2.5$k_0$), indicating reduced mode confinement or increased mode wavelength (Fig. 4e in the main text).

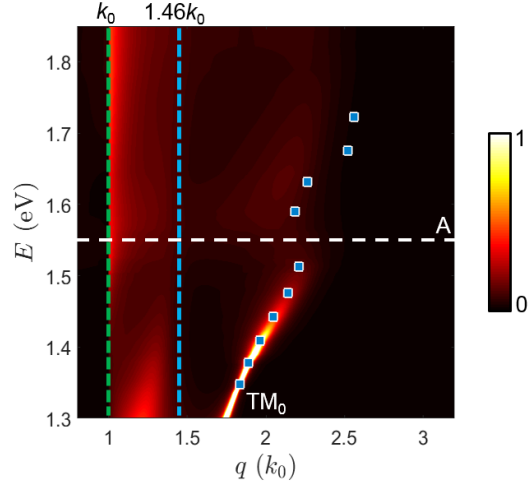

**Supplementary Figure S6.** Experimental dispersion data of waveguide EPs overlaid on the calculated dispersion color map of a 110-nm-thick MoSe$_2$ waveguide.

## 4. Effects of exciton linewidth & temperature on polariton dispersion

As discussed in the main text, the back-bending dispersion is a characteristic behavior of polaritons subject to broadening when measured by experiments under continuous-wave excitation (fixed excitation energies). Here we wish to explore the effects of the exciton broadening or linewidth on back-bending. For that purpose, we calculated the dispersion diagrams of 156-nm-thick MoSe$_2$ waveguide using a modeled *ab*-plane dielectric function with Lorentzian oscillators:

$$\varepsilon(E) = \varepsilon_1(E) + i\varepsilon_2(E) = \varepsilon_\infty + \frac{\Omega_A}{E_A^2 - E^2 - i\Gamma_A E} + \frac{\Omega_B}{E_B^2 - E^2 - i\Gamma_B E}. \quad (S5)$$

To match the experimental optical constants from literature[3] (Supplementary Fig. S4), the oscillator parameters are set to be $\varepsilon_\infty = 21$, $\Omega_A = 2 \text{ eV}^2$, $E_A = 1.55 \text{ eV}$ (A exciton energy), $\Gamma_A = 0.1 \text{ eV}$ (A exciton linewidth), $\Omega_B = 1.53 \text{ eV}^2$, $E_B = 1.85 \text{ eV}$ (B exciton energy), and $\Gamma_B = 0.12 \text{ eV}$ (B exciton linewidth).

According to Ref. 4, the exciton linewidth (broadening) in MoSe$_2$ comes from both non-thermal and thermal processes. The non-thermal processes include defect and impurity scattering or electron-electron scattering. The thermal process is mainly due to interactions with longitudinal optical phonons. The thermal contribution can be strongly suppressed at low temperature (LT). As a result, exciton linewidth at temperature less than 100 K could be less than half of that at room temperature (RT). Therefore, we adopted $\Gamma_A \approx 0.05 \text{ eV}$ as a rough estimation of the linewidths of A excitons of bulk MoSe$_2$ at low temperature, which is consistent with the experimental results given in Ref. 4.

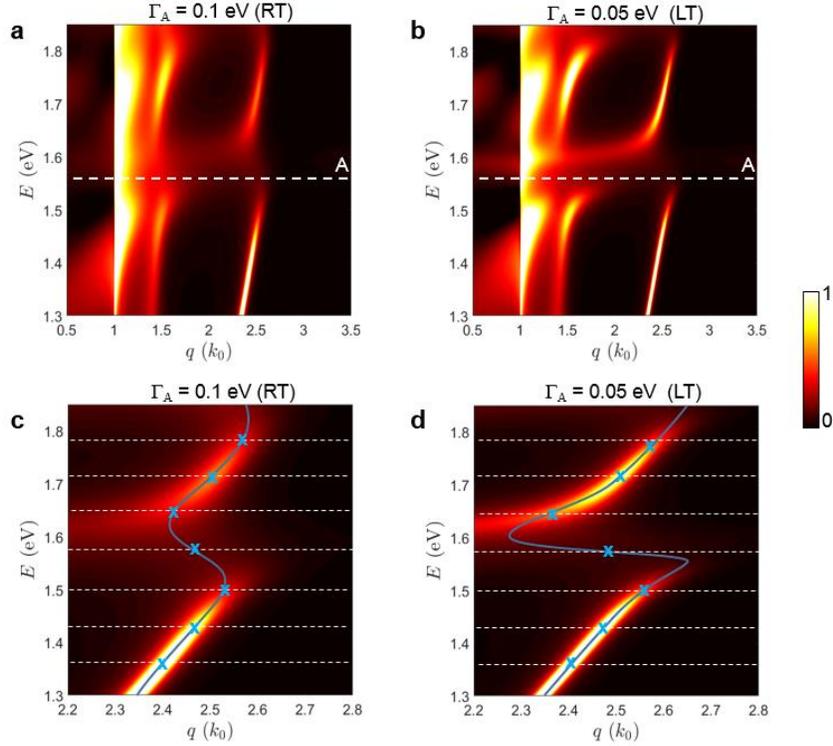

**Supplementary Figure S7. a,b**, Calculated dispersion color maps of the 156-nm-thick MoSe$_2$ waveguide with modeled *ab*-plane optical constants derived from a Lorentzian oscillator with exciton linewidths of $\Gamma_A$ =0.1 eV and 0.05 eV, corresponding to room temperature (RT) and low temperature (LT), respectively. **c,d**, Zoomed-in version of panels **a** and **b** respectively at the *q* range from 2.2$k_0$ to 2.8$k_0$. The blue crosses in **c** and **d** mark the positions with maximum photonic spectral weight along the horizontal line cuts.

The calculated dispersion diagrams with the modeled optical constants (Eq. S5) are given in Supplementary Fig. S7, where effects of exciton linewidth on polariton dispersion are clearly demonstrated. Here we compare the case of $\Gamma_A \approx 0.1$ eV (corresponding to RT, Supplementary Fig. S7a and S7c) with that of $\Gamma_A \approx 0.05$ eV (corresponding to LT, Supplementary Fig. S7b and S7d). In both cases, back-bending dispersion close to the A exciton energy is clearly seen, but the polariton mode with $\Gamma_A \approx 0.05$ eV is back-bended more dramatically, indicating stronger light-exciton coupling. In addition, the momentum broadening ($\Delta q$) of the EP mode away from the exciton energy becomes narrower at $\Gamma_A = 0.05$ eV suggesting that polaritons have smaller damping or longer propagation length.

## 5. Propagation length of the waveguide exciton polaritons

In Supplementary Fig. S8a and S8b, we plot the full-scale image and fringe profile of the 156-nm-thick MoSe$_2$ waveguide taken at *E* = 1.35 eV. Both the images (Fig. 1) and profiles (Fig. 3a) shown in the main text are truncated intentionally to fit the diagram. Here in Supplementary Fig. S8a and S8b, fringes or oscillations are seen 30 μm away from the sample edge, indicating a long propagation length ($L_p$) of these modes. In order to determine quantitatively the propagation length ($L_p$) of the measured waveguide EPs, we first extract the linewidths of the FT peaks shown in Fig. 3b of the main text. The FT linewidth data, plotted in Supplementary Fig. S9a, shows an increase with energy in the spectral range. Based on the linewidth data, we can then determine $L_p$ using Eq. S11 (see discussions below). Thus-obtained $L_p$ data points are plotted in Fig. 3c of the main text. Note that the resolution (~1/30 $\mu m^{-1}$) of the FT profiles (Fig. 3b in the main text) is limited by the sample size (~30 $\mu m$), so the linewidth data point in Supplementary Fig. S9a at the lowest photon energy ($E = 1.35$ eV) is most likely overestimated due to the resolution limit. Therefore, the extracted $L_p$ data point at this energy is possibly underestimated compared to the realistic values. In Supplementary Fig. S9b, we also plot the number of propagation cycles, namely propagate length ($L_p$) versus polariton wavelength ($\lambda_p$), at various photon energies. Here one can see that the waveguide EPs can propagate over 30 cycles before losing ~63% (1-1/$e$) of the polariton energy.

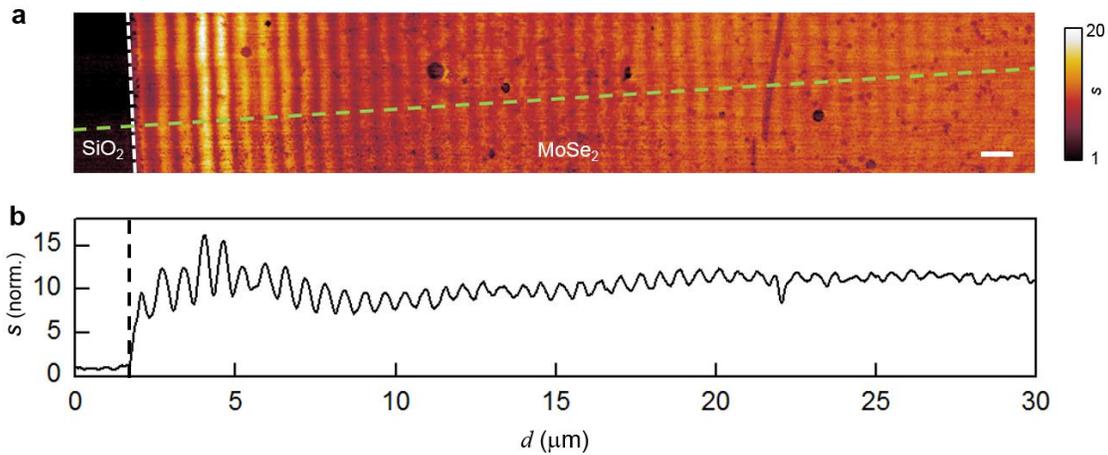

**Supplementary Figure S8**. **a**, Near-field amplitude image (*s*) of the 156-nm-thick MoSe$_2$ planar waveguide taken at $E = 1.35$ eV in the perpendicular configuration. The near-field amplitude is normalized to that of the SiO$_2$/Si substrate. The white dashed line marks the edge of the sample. The scale bar represents 1 $\mu m$. **b**, Line profile taken from **a** perpendicular to fringes along the green dashed line.

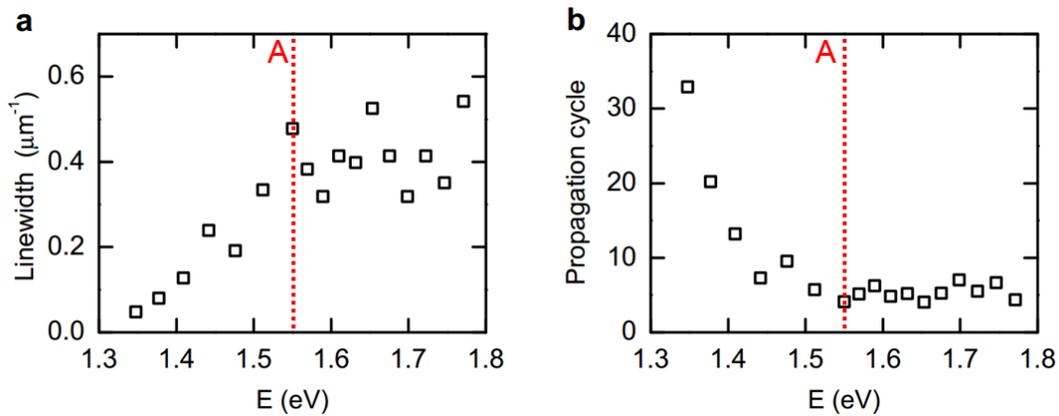

**Supplementary Figure S9**. **a**, Linewidths of the FT peaks measured from the FT profiles in Fig. 3b of the main text. **b**, Estimated number of propagation cycles ($L_p/\lambda_p$) of polaritons at various photon energies. Vertical dashed lines mark the A exciton energy.

Now we discuss how we convert the linewidths ($W$) into propagation lengths ($L_p$) in the case of perpendicular configuration. As discussed in the main text and also in Section 1 of the Supplementary Information, the total optical signal ($\mathbf{E}_{tot}$) responsible for the observed interference fringes comes from photon paths 'P1' ($\mathbf{E}_1$) and 'P2' ($\mathbf{E}_2$): $\mathbf{E}_{tot} = \mathbf{E}_1 + \mathbf{E}_2$. In path 'P1', photons are directly back-scattered by the tip to the detector. In path 'P2', photons first transfer into the waveguide EPs and then scattered back to photons by the sample edge. The momentum of the EP mode can be written as: $q_p = q_1 + iq_2$, where $q_1 = 2\pi/\lambda_p$ and $q_2 = 1/(2L_p)$. Throughout the measurement, tip is fixed and the sample is scanning, so both the amplitude and phase of $\mathbf{E}_2$ is dependent on the tip-edge distance ($d$). Therefore, we can write the amplitude of the total optical signal as:

$$|\mathbf{E}_{tot}(d)| = |\mathbf{E}_1 + \mathbf{E}_2(d)| = |A_1 + A_2(d)e^{i\Phi(d)}| \quad (x > 0). \tag{S6}$$

Here $A_1$ is the amplitude of $\mathbf{E}_1$, $A_2(d) = B_2 d^{-1/2} e^{-q_2 d}$ is the amplitude of $\mathbf{E}_2$, $\Phi(d) = [q_1 - k_0 \cos(\alpha)]d + \Phi_0$ is the relative phase difference between $\mathbf{E}_2$ and $\mathbf{E}_1$ in the perpendicular configuration (Eq. S1). For simplicity, we define $\Lambda = q_1 - k_0 \cos(\alpha)$ below. The constant $A_1$ describes the efficiency of photon back-scattering in 'P1'. The constant $B_2$ describes the conversion efficiency from photons to EPs (via tip) times the conversion efficiency from EPs to photons (via edge) in 'P2'. By substituting the formula of $A_2(d)$ and $\Phi(d)$ above into Eq. S6, we have

$$|\mathbf{E}_{tot}(d)| = \sqrt{A_1^2 + B_2^2 d^{-1} e^{-2q_2 d} + A_1 B_2 d^{-1/2} e^{-q_2 d} e^{i(\Lambda d + \Phi_0)} + A_1 B_2 d^{-1/2} e^{-q_2 d} e^{-i(\Lambda d + \Phi_0)}}. \tag{S7}$$

Considering the low conversion rates between photons and EPs, and also the exponential decaying of the EPs with $d$, we are safe to assume $A_2 = B_2 d^{-1/2} e^{-q_2 d} \ll A_1$, so Eq. S7 can be written approximately as:

$$|\mathbf{E}_{tot}(d)| \approx A_1 + \frac{1}{2} B_2 d^{-1/2} \left[ e^{i(\Lambda d + \Phi_0)} + e^{-i(\Lambda d + \Phi_0)} \right] e^{-q_2 d}. \tag{S8}$$

By performing Fourier transformation (FT) on Eq. S8, we have

$$|\mathbf{E}_{tot}(k)| \propto \left| [q_2 - i(\Lambda - k)]^{-1/2} + [q_2 + i(\Lambda + k)]^{-1/2} e^{-2\Phi_0} \right|. \tag{S9}$$

Equation S9 indicates that there should be two peaks at $k = \pm\Lambda$ when performing FT analysis of the fringe profiles. They are identical in shape and well apart from each other ($q_2 \ll \Lambda$), so we only need to consider one peak:

$$|\mathbf{E}_{tot}(k)| \propto \left[ q_2^2 + (\Lambda - k)^2 \right]^{-1/4}. \tag{S10}$$

The full width at half maximum of the peak given by Eq. S10 is $\delta k = 2\sqrt{15} q_2$. Note that the unit of the FT profiles in Figs. 2 and 3 of the main text is $1/\rho_\perp$ instead of $k = 2\pi/\rho_\perp$, so the measured linewidth ($W$) should be scaled by a factor of $2\pi$, namely $W = \sqrt{15} q_2 / \pi$. Therefore, we have:

$$L_P = (2q_2)^{-1} = \sqrt{15}/(2\pi W). \tag{S11}$$

Based on Eq. S11, we converted the measured FT linewidths (Supplementary Fig. S9a) into the propagation lengths at various excitation energies (squares in Fig. 3c).

**Supplementary References**
31. Fei, Z. et al. Gate-tuning of graphene plasmons revealed by infrared nano-imaging. *Nature* **487**,

82-85 (2012).
32. Dai, S. et al. Tunable phonon polaritons in atomically thin van der Waals crystals of boron nitride. *Science* **343**, 1125-1129 (2014).
33. Beal, A.R. & Hughes, H.P. Kramers-Kronig analysis of the reflectivity spectra of 2H-MoS$_2$, 2H-MoSe$_2$ and 2H-MoTe$_2$, *J. Phys. C: Solid State Phys.* **12**, 881-890 (1979).
34. Arora, A., Nogajewski, K., Molas, M., Koperski, M. & Potemski, M. Exciton band structure in layered MoSe$_2$: from a monolayer to the bulk limit. *Nanoscale* **7**, 20769 (2015).